\documentclass[preprint,12pt]{elsarticle}
\usepackage[nodots]{numcompress}
\usepackage{graphicx}
\usepackage{epsfig}
\usepackage{graphicx}
\usepackage[normal]{subfigure}
\usepackage{overpic}
\usepackage{float}
\usepackage{parskip}
\usepackage{natbib}
\usepackage{float}
\usepackage{dblfloatfix}
\usepackage{fixltx2e}
\usepackage{color}
\usepackage{geometry}
\usepackage{amsmath}
\usepackage{amssymb}
\usepackage{array}
\usepackage{lineno}
\usepackage[american]{babel}
\usepackage{times}
\usepackage{mathtools}
\usepackage[labelfont=bf]{caption}
\newcommand{\vmu}{\mbox{\boldmath{$\mu$}}}
\usepackage[amssymb]{SIunits}
\newcommand{\vphi}{\mbox{{\boldmath{$\phi$}}}}
\newcommand{\vc}{\mbox{{\boldmath{$c$}}}}

\journal{Acta Materialia}

\begin{document}

\begin{frontmatter}



\title{Theoretical and numerical study of lamellar eutectoid growth influenced by volume diffusion.}


\author[KIT]{Kumar Ankit\corref{cor1}}
\ead{kumar.ankit@hs-karlsruhe.de}
\author[KIT,ECOLE]{Abhik Choudhury}
\author[KIT]{Cheng Qin}
\author[KIT]{Sebastian Schulz}
\author[KIT]{Malte McDaniel}
\author[KIT]{Britta Nestler}
\cortext[cor1]{Corresponding author. Tel.:  +49 721 608-45022.}
\address[KIT]{Karlsruhe Institute of Technology (KIT), IAM-ZBS, Haid-und-Neu-Str. 7, D-76131 Karlsruhe, Germany}
\address[ECOLE]{Laboratoire PMC (Condensed Matter Physics) - Ecole Polytechnique 91128, Palaiseau Cedex, France}

\begin{abstract}
We investigate the lamellar growth of pearlite at the expense of austenite during
the eutectoid transformation in steel. To begin with, we extend the
Jackson-Hunt-type calculation (previously used to analyze eutectic transformation) to
eutectoid transformation by accounting for diffusion in all
the phases. Our principal finding is that the growth 
rates in presence of diffusion in all
the phases is different as compared to the case when diffusion in growing phases is absent.
The difference in the dynamics is described by a factor $'\rho'$ which 
comprises of the ratio of the diffusivities of the bulk and
the growing phases, along with the ratios of the slopes of the phase co-existence
lines. Thereafter, we perform phase-field simulations, the results
of which are in agreement with analytical predictions.
The phase-field simulations also reveal that 
diffusion in austenite as well
as ferrite leads to the formation 
of tapered cementite along with an overall increase in
the transformation kinetics as compared to diffusion in austenite (only).
Finally, it is worth noting that the aim of present work 
is not to consider the pearlitic transformation in totality, 
rather it is to isolate and thereby investigate the influence 
of diffusivity in the growing phases on the front velocity.
\end{abstract}


\end{frontmatter}


% %\linenumbers
\section{Introduction}
\label{intro}
The mechanism of eutectoid transformation in steel has been a topic of
theoretical as well as experimental investigation since inception of steel as a
structural material. The eutectoid transformation involves the formation of
a pearlite colony which appears as alternate lamellae of ferrite and cementite
phases, that grow as a common front with the austenite. Cementite is the carbon
rich phase whereas the carbon solubility in ferrite is relatively quite low
\cite{Zackay:1962uq,Kral:2000cr,De-Graef:2006nx}.

The two principal mechanisms of eutectoid reaction, i.e. austenite to pearlite
phase transformation, cited in the literature are volume diffusion and grain
boundary diffusion. The former suggests the volume diffusion of carbon ahead of
the phase interface, while the latter emphasizes the role of grain boundary
diffusion as the rate controlling step. The pioneering work of Zener
\cite{Zener:1947uq}, Hillert \cite{Hillert:1957vn} and Tiller
\cite{Tiller:1958zr} on pearlite formation explains the relation between the
lamellar spacing and undercooling during the phase transformation. In spite of
making a generous effort to explain the phenomenology of pearlitic
transformation, the classical Zener-Hillert model shows large deviations from
the experimentally measured lamellar growth velocities. The model assumes no
diffusion in the ferrite phase whilst considering diffusion in austenite phase
(only). This would be a reasonable assumption in case of eutectic solidification
problem, where the diffusivity in solid is lower than the diffusivity in liquid
(bulk phase) by a factor of 1000. However, in a solid state phase transformation
like the eutectoid reaction, the diffusivity in austenite (bulk phase) is
comparable to the ferrite. Thus, it is reasonable to expect
some disagreement of experimental velocities with corresponding 
values derived from the Zener-Hillert co-operative growth model. 

Jackson and Hunt \cite{Jackson:1966ly} adapt the Zener-Hillert model for
investigating directional solidification in eutectics with a constant
velocity of growth front, which broadly falls in the same class of moving
boundary problem as the eutectoid transformation.  Recently, Nakajima et. al
\cite{Nakajima:2006ve} use the multi phase-field method to simulate the
co-operative pearlite growth by accounting for diffusion in the ferrite as well
as the austenite phase. They predict a successive process of diffusion in ferrite and growth of
cementite from the ferrite, resulting in an increase of the kinetics of pearlitic transformation by
a factor of four as compared to growth from austenite exclusively. The simulated
cementite lamella is found to be tapered and exhibits a conical morphology. This
is interpreted as an effect of diffusion in ferrite. Steinbach and Plapp
\cite{Steinbach:2011ys}  claim an overlap of phase-field results with
Hillert's model in absence of diffusion in ferrite. Further, they couple a
stress-driven diffusion field to the phase field and study the effect of
transformation strains. However, Pandit and Bhadeshia \cite{Pandit:2011ab}
argue that pearlite forms by reconstructive transformation, in which case,
transformation strains should not be significant. They also emphasize the need
to consider both the mechanisms, volume as well as interfacial diffusion, 
simultaneously for an overlap with experimental findings.

In the present article, we extend our previous work on Jackson-Hunt (JH)
analysis of ternary eutectic alloys \cite{Choudhury:2011qf} to study the
eutectoid transformation. The main question which we address is: Can a JH type
analysis (previously done for eutectics) be extended to predict lamellar growth
velocities of pearlite by accounting for diffusion in austenite as well as
ferrite? In order to answer this question, we first extend the JH analysis for
eutectics by accounting for diffusion in austenite as well as ferrite.  We
analyze the case of stable lamellar coupled growth and derive the expressions
for lamellar growth velocity as a function of undercooling and lamellar spacing. 
This is followed by comparison of analytical prediction with the numerical 
results of a thermodynamically consistent phase-field model.  

The remainder of this article is organized as follows: In section~\ref{sec:Theory}
we derive an expression for lamellar growth velocity as a function of
undercooling and spacing in pearlite using a JH type analysis.  In
section~\ref{sec:PF_model}, the quantitative phase-field model used to simulate
pearlite growth is outlined. In section~\ref{sec:thermo_des}, we describe the
thermodynamic data-fitting procedure to approximate the variation of
grand-potential of the respective phases as a function of chemical potential. In
section~\ref{sec:sharp_int_lim} we derive the relation between the simulation
parameters and corresponding quantities in the sharp interface limit. In
section~\ref{sec:comparison} we compare the  lamellar growth velocity obtained
from phase-field simulations to the analytical expressions for the velocity, 
derived in section~\ref{sec:Theory}. Section~\ref{sec:conclusion} concludes the article.
\section{Theoretical analysis of coupled growth}
\label{sec:Theory}
We consider the diffusion of the components A and B ahead of the planar
eutectoid front. For calculating the concentration fields ahead of the growth
front in question, we make the following Fourier series expansion for $c_{A}$
and $c_{B}$,
\begin{eqnarray} \label{eq:Fourier Expansion_gamma}
c_{X}^\gamma = \sum_{n=-\infty}^{\infty} X_{n} e^{ik_{n}x-q_{n}z} + \left(c_{X}^\infty\right)_\gamma,
\;\;\;\;\;X= A, B
\end{eqnarray}
where $\gamma$ is the austenite phase. In the respective growing phases 
($\alpha$ and $\beta$) the concentration fields can be respectively written
as,
\begin{eqnarray} \label{eq:Fourier Expansion_nu}
c_{X}^\nu = \sum_{n=-\infty}^{\infty} X_{n} e^{ik_{n}x+q_{n}z} + \left(c_{X}^\infty\right)_\nu,
\;\;\;\;\;X= A, B\;\;\;\;\;\nu= \alpha,\beta.
\end{eqnarray}

An elaborate description of the terms involved
in the above expression and derivation from a
stationary diffusion equation has been described
in detail in the previous work
on eutectic growth \cite{Choudhury:2011qf}.
In the field under consideration,
the growth front is assumed to be at $z=0$. 
Further, $z>0$ depicts austenite phase where
exponential profiles for the concentrations of components 
A and B exist.
For $z<0$, the composition
profile in pearlite (for ferrite and cementite phases) 
have similar exponential profiles. Therefore, to account for
the symmetry across the interface, we change
the sign of the exponent $e^{-q_{n}z}$ to $e^{q_{n}z}$ when
treating the concentration profiles in ferrite and 
cementite phases ($\forall z<0$). 

In the Jackson-Hunt analysis for the calculation of diffusion field in
liquid and solid, the Stefan's condition at $\nu-\gamma$ interface, which expresses
mass-conservation upon the phase transformation reads as
\begin{align}\label{eq:Stefan Condition}
D^{\nu} \partial_{n}c^{\nu}_X|_{z=0}-D^{\gamma} \partial_{n}c^{\gamma}_X|_{z=0}= v_{n}\Delta
c_{X}^{\nu}, \;\;\;\;\;\nu= \alpha, \beta
\end{align}
where $\partial_{n}c^{\nu}_X$ denotes the partial derivative of $c^{\nu}_{X}$ in the
direction normal to the interface. The quantity $v_{n}$ is the normal velocity of the
interface (positive for a growing front) and $\Delta c_{X}^{\nu} = c_{X}^{\gamma} - c_{X}^{\nu}$. 
$D^{\gamma}$ and $D^{\nu}$ are
chemical diffusion coefficients for bulk and growing phases respectively.
For using the Stefan condition, we take the derivative of $c^{\nu}_{X}$ with respect
to the 'z' coordinate
\begin{align} \label{eq:Derivative}
\partial_{z}c_{X}^\nu\mid_{z=0} = \sum_{n=-\infty}^{\infty} q_{n} X_{n}
e^{ik_{n}x}\;\;\;\;\; \nu=\alpha,\beta
\end{align}
for the growing phases and 
\begin{align} \label{eq:Derivative}
\partial_{z}c_{X}^\gamma\mid_{z=0} = \sum_{n=-\infty}^{\infty} -q_{n} X_{n}
e^{ik_{n}x},
\end{align}
for the austenite phase. Integration across one 
lamella period (lamellar spacing) $\lambda$ 
gives,
\begin{eqnarray} \label{eq:Integral}
q_{n}X_{n}^{\alpha}D^{\alpha}\eta_{\alpha}\lambda+q_{n}X_{n}^{\beta}D^{\beta}
\eta_{\beta}\lambda+q_{n}X_{n}^{\gamma}D^{\gamma}\lambda =
\sum_{j=0}^{M-1}\int_{x_{j}\lambda}^{x_{j+1}\lambda} v_{n}\Delta c_{X}^{\nu_{j}}e^{-ik_{n}x}dx.
\end{eqnarray}
where, M denotes the number of lamellae
and $\eta_{\alpha}$ and $\eta_{\beta}$ 
are the respective phase volume fractions.
In the above equation, there are three sets of 
unknowns $X_n^\alpha$, $X_n^\beta$ and 
$X_n^\gamma$ 
in contrast to the classical Jackson-Hunt 
type analysis with vanishing diffusivity in
the solid, where $X_n^\gamma$ remains the
only unknown, and is thereby fixed by
the property of orthogonality of the
respective modes. In the present situation, 
we need a relation among the fourier coefficients,
to fix the respective unknowns. We achieve 
these relations by arguing that 
the constitutional undercooling can be
derived equivalently by using either
the shifts in the equilibrium concentrations
of the bulk or the growing phases. Hence,
the resultant shift in the average concentration
in the $\gamma$ phase and the corresponding 
shift in the $\alpha$ and $\beta$ phases
must be constrained by the relation,
\begin{align}
m^\alpha_{X}\big(\langle c^\alpha_X \rangle - c_{X,E}^\alpha\big)&=
m^{\alpha,\gamma}_{X}\big(\langle c^\gamma_X \rangle - c_{X,E}^\gamma\big)\\
m^\beta_{X}\big(\langle c^\beta_X \rangle - c_{X,E}^\beta\big)&=
m^{\beta,\gamma}_{X}\big(\langle c^\gamma_X \rangle - c_{X,E}^\gamma\big)
\end{align}

where $\langle c^{\alpha}_{X}\rangle$, 
$\langle c^{\beta}_{X}\rangle$ 
and $\langle c^{\gamma}_{X}\rangle$ 
denote the average phase concentrations
at the interface, while
$c^{\alpha}_{X,E}$, $c^{\beta}_{X,E}$ and $c^{\gamma}_{X,E}$ are 
the eutectoid compositions. 
$m^{\alpha}_{X}$ and $m^{\beta}_{X}$ represent the slopes
(with respect to the concentration of component 'X')
of the $\alpha$ and $\beta$ phase in equilibrium
with austenite. Similarly,  
$m^{\alpha,\gamma}_{X}$ and $m^{\beta,\gamma}_{X}$ 
represent the slopes of the co-existence
lines of the austenite phase in equilibrium
with $\alpha$ and $\beta$ phases respectively.

While such an equation certainly has 
multiple solutions, we invoke the 
following assumption,
\begin{eqnarray} \label{eq:Condition}
m^{\alpha}_{X}X_{n}^{\alpha} = m^{\alpha,\gamma}_{X}X_{n}^{\gamma} 
\\ m^{\beta}_{X}X_{n}^{\beta} = m^{\beta,\gamma}_{X}X_{n}^{\gamma} 
\end{eqnarray}
which satisfies the given property.
Substituting the preceding condition in equation \ref{eq:Integral} yields,
\begin{eqnarray}
\\ q_{n}X_{n}^{\gamma} \delta_{nm}\lambda\left[ \dfrac{D^{\alpha}}{D^{\gamma}}
\dfrac{m^{\alpha,\gamma}_{X}}{m^{\alpha}_{X}}\eta_{\alpha} + \dfrac{D^{\beta}}{D^{\gamma}}
\dfrac{m^{\beta,\gamma}_{X}}{m^{\beta}_{X}}\eta_{\beta} + 1\right] =
\sum_{j=0}^{M-1}\int_{x_{j}\lambda}^{x_{j+1}\lambda}\dfrac{v_{n}}{D^{\gamma}}
\Delta c_{X}^{\nu_{j}} e^{-ik_{n}x} dx\nonumber
\\ = \dfrac{2}{l}\sum_{j=0}^{M-1}\int_{x_{j}\lambda}^{x_{j+1}\lambda}
e^{-ik_{n}x}\Delta c_{X}^{\nu_{j}} dx \nonumber
\end{eqnarray}
and hence, we rearrange to
\begin{align}
X_{n}^{\gamma}=\dfrac{4}{lq_{n}\lambda k_{n}\left[
\dfrac{D^{\alpha}}{D^{\gamma}} \dfrac{m^{\alpha,\gamma}_{X}}{m^{\alpha}_{X}}\eta_{\alpha} +
\dfrac{D^{\beta}}{D^{\gamma}} \dfrac{m^{\beta,\gamma}_{X}}{m^{\beta}_{X}}\eta_{\beta} +
1\right]}\sum_{j=0}^{M-1} \Delta c_{X}^{\nu_{j}}e^{-ik_{n}\lambda\left(
x_{j+1}+x_{j}\right)/2 } \sin\left[ k_{n}\lambda\left(
x_{j+1}-x_{j}\right)/2\right] 
\end{align}
where $\nu_{j}$ represents the name of one of the growing phases
($\alpha,\beta$) occurring in the sequence of M lamellae 
($\nu_{0}, \nu_{1}, \nu_{2}\dots\nu_{M-1}$) periodically 
arranged with a repeat distance (lamellar spacing) $\lambda$.
The symbol $l$ appearing in the denominator represents the 
characteristic length scale of the concentration boundary layer.
By considering the negative summation indices, we can formulate real
and imaginary combinations of these coefficients,
\begin{eqnarray}
X_{n}^{\gamma}+X_{-n}^{\gamma}=\dfrac{8}{lq_{n}\lambda k_{n}\rho}
\sum_{j=0}^{M-1} \Delta c_{X}^{\nu_{j}}\cos\left[ k_{n}\lambda\left(
x_{j+1}+x_{j}\right)/2\right]\sin\left[ k_{n}\lambda\left(
x_{j+1}-x_{j}\right)/2\right] \nonumber
\\ i\left( X_{n}^{\gamma}-X_{-n}^{\gamma}\right) =\dfrac{8}{lq_{n}\lambda
k_{n}\rho} \sum_{j=0}^{M-1} \Delta c_{X}^{\nu_{j}}\sin\left[ k_{n}\lambda\left(
x_{j+1}+x_{j}\right)/2\right]\sin\left[ k_{n}\lambda\left(
x_{j+1}-x_{j}\right)/2\right] 
\end{eqnarray}
where, 
\begin{eqnarray}
\rho = \dfrac{D^{\alpha}}{D^{\gamma}} \dfrac{m^{\alpha,\gamma}_{X}}{m^{\alpha}_{X}}\eta_{\alpha} +
\dfrac{D^{\beta}}{D^{\gamma}} \dfrac{m^{\beta,\gamma}_{X}}{m^{\beta}_{X}}\eta_{\beta} + 1 
\end{eqnarray}
Therefore, equation \ref{eq:Fourier Expansion_gamma} can be rewritten as,
\begin{align}
c_{X}^\gamma=\left(c_{X}^{\infty}\right)_{\gamma} +
\dfrac{X_{0}^\gamma}{\rho}+\dfrac{1}{\rho}\sum_{j=0}^{M-1}\sum_{n=1}^{\infty}\dfrac{8}{
lq_{n}\lambda k_{n}}\cos\left[ k_{n}\lambda\left(
x_{j+1}+x_{j}\right)/2\right]\sin\left[ k_{n}\lambda\left(
x_{j+1}-x_{j}\right)/2\right]\cos\left( {k_{n}x}\right)  \nonumber
\\ +\dfrac{1}{\rho}\sum_{j=0}^{M-1}\sum_{n=1}^{\infty}\dfrac{8}{lq_{n}\lambda
k_{n}}\sin\left[ k_{n}\lambda\left( x_{j+1}+x_{j}\right)/2\right]\sin\left[
k_{n}\lambda\left( x_{j+1}-x_{j}\right)/2\right]\sin\left( {k_{n}x}\right) 
\end{align}
The general expression for the mean concentration $\langle c_{X}^\gamma\rangle_{m}$
ahead of the $m$th phase of the phase sequence can be calculated to yield
\begin{eqnarray}\label{eq:mean_concentration}
\begin{split}
\langle c_{X}^\gamma\rangle_{m} =& \dfrac{1}{\left(
x_{m+1}-x_{m}\right)\lambda}\int_{x_{m}\lambda}^{x_{m+1}\lambda}c_{X}^\gamma dx 
\\ =& \left(c_{X}^{\infty}\right)_\gamma + \dfrac{X_{0}^\gamma}{\rho} +\dfrac{1}{\left(
x_{m+1}-x_{m}\right)\rho}\sum_{n=1}^{\infty}\sum_{j=0}^{M-1}\left\lbrace
\dfrac{16}{\lambda^{2}k_{n}^{2}lq_{n}}\Delta c_{X}^{\nu_{j}} \right. 
 \\ &\left.\times\sin\left[\pi n\left(x_{m+1}-x_{m}\right)\right]\sin\left[\pi
n\left(x_{j+1}-x_{j}\right)\right]\times\cos\left[\pi
n\left(x_{m+1}+x_{m}-x_{j+1}-x_{j}\right)\right]\vphantom{\dfrac{16}{\lambda^{2}
k_{n}^{2}lq_{n}}}\right\rbrace
\end{split}
\end{eqnarray}

For the binary eutectoid system with phases $\alpha,\ \beta, \mathrm{and} \ \gamma $,
we can derive the average concentrations of the components A, B by setting, $x_{0} =0,\
x_{1}=\eta_{\alpha},\ x_{2}=1$ and applying equation \ref{eq:mean_concentration}
\begin{align}
\langle c_{X}^\gamma\rangle_{\alpha}=& \left(c_{X}^{\infty}\right)_\gamma +
\dfrac{X_{0}^\gamma}{\rho}+\dfrac{1}{\eta_{\alpha}\rho}\sum_{n=1}^{\infty}\left\lbrace
\dfrac{16}{\lambda^{2}k_{n}^{2}lq_{n}}\left(\Delta c_{X}^{\alpha}-\Delta
c_{X}^{\beta}\right)\times\sin^{2}\left(\pi n\eta_{\alpha}\right)\right\rbrace\nonumber
\\ \cong & \left(c_{X}^{\infty}\right)_\gamma +
\dfrac{X_{0}^\gamma}{\rho}+\dfrac{2\lambda}{\eta_{\alpha}\rho l}{\cal
P}\left(\eta_{\alpha}\right)\Delta c_{X}
\\ \langle c_{X}^\gamma\rangle_{\beta}=& \left(c_{X}^{\infty}\right)_ \gamma+
\dfrac{X_{0}^\gamma}{\rho}-\dfrac{2\lambda}{\left(1-\eta_{\alpha}\right)\rho l}{\cal
P}\left(1-\eta_{\alpha}\right)\Delta c_{X}
\end{align}
with $k_{n}=2\pi n/\lambda$, $q_{n}\approx k_{n}$, $\lambda / l \ll 1$, $\Delta
c_{X}= \Delta c_{X}^{\alpha}-\Delta c_{X}^{\beta}$ and the dimensionless function
\begin{align}
{\cal P}\left(\eta\right)=\sum_{n=1}^{\infty}\dfrac{1}{\left(\pi
n\right)^{3}}\sin^{2}\left(\pi n\eta\right)
\end{align}
which has the property ${\cal P}\left(\eta\right) = {\cal
P}\left(1-\eta\right) = {\cal P}\left(\eta-1\right)$

Incorporating the Gibbs-Thomson effect and using the relation $l = 2D^{\gamma}/v$ leads to,
\begin{align}
\Delta T_{\alpha} = -m_{B}^{\alpha,\gamma}B_{0}^\gamma-\dfrac{\lambda v}{\eta_{\alpha}D^{\gamma} \rho}{\cal
P}\left(\eta_{\alpha}\right)m_{B}^{\alpha,\gamma}\Delta c_{B} +
\Gamma_{\alpha}\langle\kappa\rangle_{\alpha}
\\ \Delta T_{\beta} = -m_{A}^{\beta,\gamma}A_{0}^\gamma-\dfrac{\lambda v}{\eta_{\beta}D^{\gamma}\rho}{\cal
P}\left(\eta_{\beta}\right)m_{A}^{\beta,\gamma}\Delta c_{A} +
\Gamma_{\beta}\langle\kappa\rangle_{\beta}
\end{align}
where $\langle\kappa\rangle_{\alpha} = 2\sin\theta_{\alpha\beta}/
\left(\eta_{\alpha}\lambda\right)$, $\langle\kappa\rangle_{\beta} =
2\sin\theta_{\beta\alpha}/ \left(\eta_{\beta}\lambda\right)$, $\Gamma_{\alpha}=
\tilde{\sigma}_{\alpha\gamma}T_{E}/L_{\alpha}$ and $\Gamma_{\beta}=
\tilde{\sigma}_{\beta\gamma}T_{E}/L_{\beta}$. Additionally, for a binary alloy,
the coefficients follow the condition, $B_{0}^\gamma = -A_{0}^\gamma$. 
The global front undercooling is determined using the
assumption of equal interface undercooling $\Delta T_{\alpha}=\Delta T_{\beta} =
\Delta T$. For a constant undercooling, we deduce
the relation between the growth velocity  `v' and lamellar width `$\lambda$' by
eliminating the unknown amplitude $A_{0}^{\gamma} \left(\mathrm{or} B_{0}^{\gamma}\right)$ as,
\begin{align}
v = \dfrac{\Delta
T-\dfrac{2T_{E}}{\lambda\left(m_{A}^{\beta,\gamma}+m_{B}^{\alpha,\gamma}\right)}\left[\dfrac{
m_{B}^{\alpha,\gamma}\tilde{\sigma}_{\beta\gamma}\sin\theta_{\beta\alpha}}{\eta_{\beta}
L_{\beta}}+\dfrac{m_{A}^{\beta,\gamma}\tilde{\sigma}_{\alpha\gamma}\sin\theta_{
\alpha\beta}}{\eta_{\alpha}L_{\alpha}}\right]}{-\dfrac{\lambda}{D^{\gamma}\rho}\left(\dfrac{
m_{A}^{\beta,\gamma}m_{B}^{\alpha,\gamma}}{m_{A}^{\beta,\gamma}+m_{B}^{\alpha,\gamma}}\right)\left[\dfrac{{
\cal P}\left(\eta_{\alpha}\right)\Delta c_{B}}{\eta_{\alpha}}+\dfrac{{\cal
P}\left(\eta_{\beta}\right)\Delta c_{A}}{\eta_{\beta}}\right]}.
 \label{eq:vel_lambda}
\end{align}
It is worth clarifying that in the derived expression for 
binary eutectoids above, the growing phases $\alpha$ and $\beta$
posses slopes with opposite signs for the same component. 
$m_{A}^{\beta,\gamma}$ and
$m_{B}^{\alpha,\gamma}$ denote the slopes 
of the phases $\beta$ and $\alpha$ respectively
with respect to the minority component in
each phase. Hence, they will always be of the same sign.
Therefore, for 
eutectoid systems, the denominator 
$m_{A}^{\beta,\gamma}+m_{B}^{\alpha,\gamma}$,
is non-vanishing.
Further, it is important to point the difference 
of the preceding expression with respect to
the relations for the velocity derived for
eutectic solidification in absence of 
diffusion in solid-phases 
\cite{Jackson:1966ly,Choudhury:2011qf}.
The velocity differs by the factor $\rho$,
which depends on the ratio of the
diffusivities in the growing phases
and the bulk phase, together with 
the ratio of the slopes of the
respective phase-coexistence
lines. 
 
\section{Phase-field model}
\label{sec:PF_model}
In the following investigation, we study phase evolution in a ternary
system using a quantitative phase-field model \cite{Choudhury:2012uq}. We start
by writing down the grand-potential functional of the system,
incorporating the interfacial and bulk contributions of the
respective phases. The evolution equations for the phase and 
concentration fields can be evaluated
in the standard way. Phase evolution is determined by the phenomenological
minimization of the modified functional which is formulated as the {\em grand
potential functional},
\begin{eqnarray}
 {\Omega}\left(T,\vmu,\vphi\right)&=&\int_{V}\left(\Psi\left(T,\vmu,
\vphi\right)+\left(\epsilon \tilde{a}\left(\vphi,\nabla \vphi\right) +
\dfrac{1}{\epsilon}\tilde{w}\left(\vphi\right)\right)\right)dV.
 \label{GrandPotentialfunctional}
\end{eqnarray}
We write the grand potential density $\Psi$, as an interpolation of the
individual grand potential densities $\Psi_\alpha$, where $\Psi_\alpha$ are
functions of the chemical potential $\vmu$ and temperature T in the system,
\begin{eqnarray}
 \Psi\left(T,\vmu,\vphi\right) &=&
\sum_{\alpha=1}^N\Psi_{\alpha}\left(T,\vmu\right)h_\alpha\left(\vphi\right)
\quad \textrm{with},
\label{GP_interpolation}\\
\Psi_{\alpha}\left(T,\vmu\right) &=&
f_\alpha\left(\vc^\alpha\left(\vmu\right),T\right) - \sum_{i=1}^{K-1}\mu_{i}
c_{i}^{\alpha}\left(\vmu, T\right)
\end{eqnarray}
The concentration $c_{i}^{\alpha}\left(\vmu, T\right)$ 
is an inverse of the function $\mu_{i}^{\alpha}\left(\vc, T\right)$ for every
phase $\alpha$ and component $i$. From equation \ref{GP_interpolation}, the
following relation can be derived,
\begin{eqnarray}
 \dfrac{\partial \Psi\left(T,\vmu,\vphi\right)}{\partial \mu_{i}} &=&
\sum_{\alpha=1}^N\dfrac{\partial \Psi_\alpha\left(T,\vmu\right)}{\partial
\mu_{i}}h_{\alpha}\left(\vphi\right).
\end{eqnarray}
where, 
$h_{\alpha}\left(\phi\right) = \phi_{\alpha}^{2}\left(3-2\phi_{\alpha}\right) 
+ 2\phi_\alpha\sum_{\beta<\gamma, (\beta,\gamma) \neq \alpha}\phi_\beta \phi_\gamma$.
Since, the grand potential density $\Psi\left(T,\vmu,\vphi\right)$, is the
\emph{Legendre transform } of the free energy density of the system $f\left(T,
\vc, \vphi\right)$, and from their coupled relation $\dfrac{\partial
\Psi\left(T,\vmu,\vphi\right)}{\partial \mu_i}=-c_i$, it follows that, 
\begin{eqnarray}
c_{i} &=& \sum_{\alpha=1}^N c_{i}^\alpha\left(\vmu,
T\right)h_\alpha\left(\vphi\right).
\label{conc_constraint}
\end{eqnarray}
The evolution equation for the N phase-field variables can be written as,
\begin{align}
\tau \epsilon \dfrac{\partial \phi_{\alpha}}{\partial t}=\epsilon \left(\nabla
\cdot \dfrac{\partial \tilde{a}\left(\vphi,\nabla \vphi\right)}{\partial \nabla
\phi_{\alpha}}- \dfrac{\partial \tilde{a}\left(\vphi,\nabla
\vphi\right)}{\partial \phi_\alpha}\right)-\dfrac{1}{\epsilon}\dfrac{\partial
\tilde{w}\left(\vphi\right)}{\partial \phi_\alpha}-\dfrac{\partial
\Psi\left(T,\vmu, \vphi\right)}{\partial \phi_\alpha}- \Lambda,
\label{Equation6_grandchem}
\end{align}
where $\Lambda$ is the Lagrange parameter to maintain the constraint
$\sum_{\alpha=1}^N \phi_\alpha =1$. $\tilde{a}\left(\vphi,\nabla \vphi\right)$
represents the gradient energy density and has the form,
\begin{align}
 \tilde{a}\left(\vphi,\nabla \vphi\right)=
\begin{array}{ll}
 \displaystyle \sum_{\substack{
\alpha, \beta = 1 \\
(\alpha < \beta)}}^{N,N}
\tilde{\sigma}_{\alpha \beta}\lvert q_{\alpha \beta}\lvert^{2},
\end{array}
\end{align}
where $q_{\alpha \beta}=\left(\phi_{\alpha}\nabla
\phi_{\beta}-\phi_{\beta}\nabla \phi_{\alpha}\right)$ is a 
normal vector to the $\alpha-\beta$ interface.
The double obstacle potential $\tilde{w}\left(\vphi\right)$ which is 
previously described in \cite{Nestler:2005uq,Steinbach:1999fk,Stinner:2004fk} 
can be written as,
\begin{align}
\tilde{w}\left(\vphi\right)=
\begin{array}{ll}
\dfrac{16}{\pi^{2}}\displaystyle \sum_{\substack{
\alpha, \beta = 1 \\
(\alpha < \beta)}}^{N, N}
\tilde{\sigma}_{\alpha \beta}\phi_{\alpha}\phi_{\beta},
\end{array}
\end{align}
where $\tilde{\sigma}_{\alpha \beta}$ is the surface energy. The parameter
$\tau$ is written as $\dfrac{\sum_{\alpha<\beta}^{N,N}
\tau_{\alpha\beta}\phi_\alpha\phi_\beta}{\sum_{\alpha<\beta}^{N,N}
\phi_\alpha\phi_\beta}$, where $\tau_{\alpha\beta}$ is the relaxation constant
of the $\alpha-\beta$ interface.

The concentration fields are obtained by a mass conservation equation for each
of the $K-1$ independent concentration variables $c_i$. The evolution equation 
for the concentration fields can be derived as,
\begin{eqnarray}
 \dfrac{\partial c_i}{\partial t} &=& \nabla \cdot
\left(\sum_{j=1}^{K-1}M_{ij}\left(\vphi\right) \nabla \mu_j\right).
\label{Concentration_equation}
\end{eqnarray}
\begin{eqnarray}
 M_{ij}\left(\vphi\right) &=& \sum_{\alpha=1}^{N}M_{ij}^\alpha
g_{\alpha}\left(\vphi\right),
\end{eqnarray}
where each $M_{ij}^\alpha$ represents the mobility matrix, 
of the phase $\alpha$, calculated by 
multiplying the diffusivity matrix with susceptibility matrix as,
\begin{eqnarray}
 M_{ij}^\alpha &=& D^\alpha_{ik} \dfrac{\partial
c_k^\alpha\left(\vmu,T\right)}{\partial \mu_j}.
\end{eqnarray}

In the above expression (written in Einstein notation
for a shorter description), a repeated index
implies sum over all
the elements. Every $M_{ij}^\alpha$ value is
weighed with respect to the phase 
fractions represented by $g_\alpha\left(\vphi\right)$
which gives the total mobility $M_{ij}\left(\vphi\right)$.
The function $g_\alpha\left(\vphi\right)$ is in
general not same as $h_\alpha\left(\vphi\right)$ which interpolates the grand
potentials, however, in the present description,
we utilize the same. $D_{ij}^\alpha$ represent the inter-diffusivities in phase $\alpha$
and so on.
Both the evolution equations require the information about the chemical
potential $\vmu$. Two possibilities exist to determine the unknown chemical
potential $\vmu$.
\begin{itemize}
\item The chemical potential $\vmu$ can be derived from the constraint relation
in equation \ref{conc_constraint}.
The $K-1$ independent components $\mu_i$ are  determined by simultaneously
solving the $K-1$ constraints for each of the 
$K-1$ independent concentration variables $c_i$, from the given values of $c_i$
and $\phi_\alpha$ at a given grid point. A Newton iteration scheme can be used
for solving the system of equations,
\begin{align}
 \left\lbrace\mu^{n+1}_i\right\rbrace &= \nonumber\\
& \left\lbrace\mu^{n}_i\right\rbrace -
\left[\sum_{\alpha=1}^N 
h_\alpha\left(\vphi\right)\dfrac{\partial
c_i^\alpha\left(\vmu^n,T\right)}{\partial \mu_j}\right]_{ij}^{-1}\left\lbrace
c_i-\sum_{\alpha=1}^N
c_i^\alpha\left(\vmu^n,T\right)h_\alpha\left(\vphi\right)\right\rbrace,
\label{Mu_implicit}
\end{align}
where $\left\lbrace\right\rbrace$ represents a vector 
while $\left[ \right]$ denotes a matrix.
\item Alternatively,  explicit evolution equations
for all the $K-1$ independent chemical potentials,
can be formulated by inserting the constraint equation 
\ref{conc_constraint}
into the evolution equation for each of the
concentration fields. For a general, multi-phase, 
multi-component system, 
the evolution equations for the components of the chemical potential
$\vmu$ can be written in matrix form by,
\begin{align}
 &\left\lbrace\dfrac{\partial \mu_i}{\partial t}\right\rbrace = \nonumber\\
&\left[\sum_{\alpha=1}^N 
h_\alpha\left(\vphi\right)\dfrac{\partial c_i^\alpha\left(\vmu,
T\right)}{\partial
\mu_j}\right]^{-1}_{ij}\left\lbrace\nabla\cdot\sum_{j=1}^{K-1}M_{ij}
\left(\vphi\right)\nabla\mu_j 
- \sum_{\alpha=1}^N c^\alpha_{i}\left(\vmu,T\right)\dfrac{\partial
h_\alpha\left(\vphi\right)}{\partial t}\right\rbrace.
\label{Mu_explicit}
\end{align}
\end{itemize}
In the present work, an explicit formulation,
as shown in equation \ref{Mu_explicit}
have been used  for calculating $K-1$ independent 
chemical potentials.
\section{Thermodynamic description}
\label{sec:thermo_des}
In order to describe the thermodynamics
of the respective phases, we approximate the 
variation of the grand-potential of the
respective phases using a polynomial of
second degree in the chemical potential.
Without going into the details, we would like 
to state that such an approximation
is the minimum requirement for fitting
the Gibbs-Thomson coefficients of the
respective interfaces. In our present
investigation we are especially interested
in the coupled growth of ferrite($\alpha$) 
and cementite($\beta$)
in austenite($\gamma$). Hence, the Gibbs-Thomson 
coefficients of the $\alpha-\gamma$ and
$\beta-\gamma$ interfaces are 
important parameters required for the
correct description of the system. 
Also, since we are treating a binary
system, we have only a single
independent chemical potential
$\mu$, and we define our functions
with only this chemical potential
as the argument.
We start by writing the grand-potential
of a given phase as,
\begin{align}
 \Psi^\alpha\left(T,\mu\right) &= A^\alpha\left(T\right)\mu^2 +
B^\alpha\left(T\right)\mu + C^\alpha\left(T\right).
\end{align}
At the eutectoid temperature $T_E$, 
where the three phases($\alpha,\beta,\gamma$) 
are at equilibrium, we fix the 
coefficients in the 
following manner,
\begin{align}
 A^\alpha\left(T_E\right) &=
\dfrac{1}{2}\left(\dfrac{\partial^2\Psi^\alpha}{\partial
\mu^2}\right)_{T_E,\mu_{eq}}\equiv\dfrac{-V_m}
{\dfrac{\partial^2G^\alpha}{\partial c^2}}\\
B^\alpha\left(T_E\right) &= -c -2A^\alpha\dfrac{\mu_{eq}}{V_m}\\
C^\alpha\left(T_E\right) &= \dfrac{G^\alpha}{V_m} +
A^\alpha\left(T_E\right)\left(\dfrac{\mu_{eq}}{V_m}\right)^2
\label{Fix_coefficients}
\end{align}

where $V_m$ is the molar volume, $G^\alpha$ is
the free energy of the phase at the eutectoid 
composition and the respective equilibrium temperature.
The first derivative 
$\dfrac{\partial G^\alpha}{\partial c}$, which 
is the chemical potential $\mu_{eq}$ and the
second derivative of the free energy $\dfrac{\partial ^2 G^\alpha}{\partial
c^2}$, 
are also extracted from the thermodynamic functions
of the respective phases derived from the CALPHAD
database \cite{Gustafson:1985fk}. 
We point out that while 
the above procedure fixes the
coefficients, A, B, C for austenite
and ferrite, the calculation of corresponding 
values for the cementite phase, are not as 
generic. Theoretically, 
the free energy as a function of composition
of the cementite phase is represented by a point (only). 
However, for phase-field calculations,
we require information about the changes
in free energy, at least for small differences in composition.
The database we used for this calculation yields 
two free energy values corresponding to 
compositions on either side of the eutectoid composition
(of cementite). We utilize this information 
together with the chemical potential of the cementite
phase in equilibrium with the austenite
(at the eutectoid composition) to derive
the free energy as a function
of composition for the cementite phase 
(a parabola with a narrow opening). 
The grand-potential description is 
obtained by performing the Legendre
transform of the free energy density
which gives the grand-potential density 
(a wide-inverted parabola).
It is worth mentioning
that a  limiting case (not applied here) 
could also be adopted
for stoichiometric compounds (in general)
by assuming the coefficient $A$ of cementite 
to be zero. This is equivalent to assuming 
$\dfrac{\partial^2 f}{\partial c^2}$ as infinity.
The coefficients $B$ and $C$ can thereafter
be fixed using the information about the
equilibrium chemical potential and the 
grand-potential both of which are thermodynamically
defined for the stoichiometric compound as well.

To fit the information of the slopes of 
the equilibrium lines of the phase diagram, 
it is essential to describe the variation of
the grand-potential as a function of temperature.
We achieve this through a linear 
interpolation of the coefficients of
the grand-potential density functions.
The properties at a chosen lower
temperature of ferrite and austenite
are fixed by utilizing the
same procedure as in equation \ref{Fix_coefficients},
however using the values of the
free energy $G^\alpha$, the chemical
potential and the second derivative
of the free energy with respect to 
concentration are derived from the thermodynamic
functions in CALPHAD at the chosen
temperature, and are computed at the
respective phase concentrations at
the eutectoid temperature.
Due to unavailability of the chemical 
potential for cementite from the 
CALPHAD databases,
the case of the cementite is
treated differently.
We derive this information,
using the equilibrium between the
cementite and austenite and
approximating the chemical potential
and grand-potential 
from the value given for the austenite
at the chosen temperature, as
both these quantities must be equal for
the two phases at equilibrium. The
procedure suffices because the 
austenite-cementite equilibrium is relevant
during coupled growth of the ferrite
and the cementite phases from the
austenite.

It is noteworthy that in the
present work, we do not model the difference
in diffusive mechanisms of the cementite phase.
While it may not be completely appropriate for the case of eutectoid
reaction in steels, yet it does not hamper 
the spirit of the present work which 
is the comparison between phase-field simulations
and a modified Jackson-Hunt theory. Therefore, 
in the present discussion, the set of parameters
chosen for the diffusion of carbon in cementite are exemplary
only for a possible eutectoid transformation. 

\begin{figure}
\begin{center}
\includegraphics[scale=0.9]{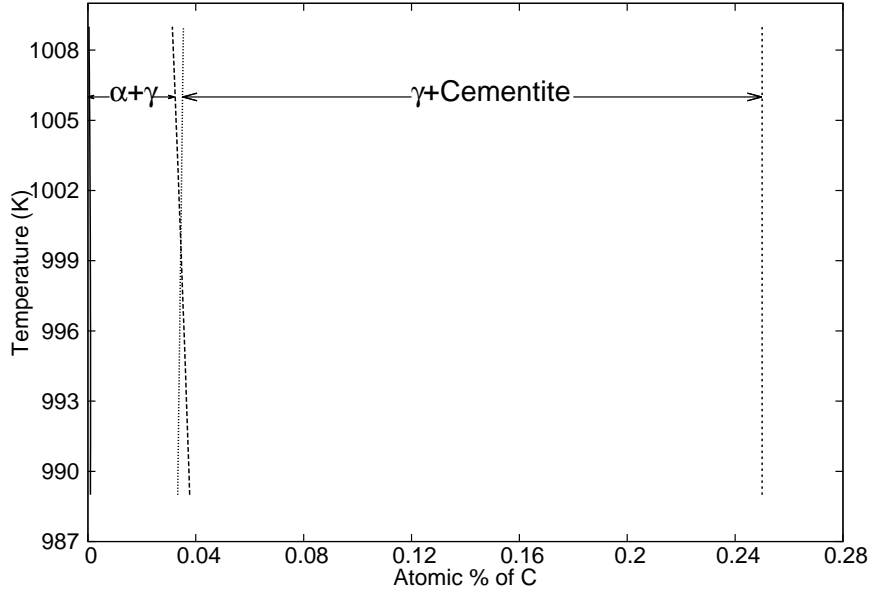}
\caption{Calculated phase diagram}
\label{fig:Phase_Diag}
\end{center}
\end{figure}
\section{Relation to sharp-interface limit}
\label{sec:sharp_int_lim}
In this section we derive the 
relation between the simulation
parameters and the corresponding
quantities in the sharp interface 
limit. Two of these quantities are 
the Gibbs-Thomson coefficient
and the slopes of the equilibrium 
co-existing lines. These relations
can be derived from the Clausius-Clapeyron
equation for the case of binary alloys 
which writes as,
\begin{align}
 \dfrac{\partial \mu}{\partial T} &= \dfrac{\left(\dfrac{\partial
\Psi^\alpha}{\partial T}-
\dfrac{\partial \Psi^\gamma}{\partial
T}\right)}{\left(c^\alpha-c^\gamma\right)}.
\end{align}
Expanding $\dfrac{\partial \mu}{\partial T}=
\dfrac{\partial \mu^{\alpha,\gamma}}{\partial c}\dfrac{\partial
c^{\alpha,\gamma}}{\partial T}$
we derive the slope of the equilibrium 
co-existence lines as,
\begin{align}
 m^{\alpha,\gamma} &= \dfrac{\partial c^{\alpha,\gamma}}{\partial
\mu}\dfrac{\left(c^\alpha-c^\gamma\right)}{\left(\dfrac{\partial
\Psi^\alpha}{\partial T}-
\dfrac{\partial \Psi^\gamma}{\partial
T}\right)}=\dfrac{-1}{2A^{\alpha,\gamma}}\dfrac{\left(c^\alpha-c^\gamma\right)}{
\left(\dfrac{\partial \Psi^\alpha}{\partial T}-
\dfrac{\partial \Psi^\gamma}{\partial T}\right)}.
\end{align}

Similarly the Gibbs-Thomson coefficient $\Gamma_{\alpha \gamma}$
derives as;
\begin{align}
 \Gamma_{\alpha \gamma} &= \dfrac{\tilde{\sigma}_{\alpha \gamma}}{\left(\dfrac{\partial
\Psi^\alpha}{\partial T}-
\dfrac{\partial \Psi^\gamma}{\partial T}\right)},
\end{align}

where $\tilde{\sigma}_{\alpha \gamma}$ is 
the surface tension of the 
$\alpha-\gamma$ interface.
Similar expressions for
the Gibbs-Thomson coefficients of the
cementite phase can be derived by replacing
$\alpha$ with $\beta$ in the preceding
equations. Theoretically,
the deviations of chemical potential or the concentrations
due to curvature, would be non-existent for the
cementite phase, by the virtue of 
the infinite values of $\dfrac{\partial ^2f }{\partial c^2}$.
However, the approximate construction of the 
grand-potential densities allow small deviations in
composition due to curvature effects,
although of a much lower order as compared
to the austenite and ferrite phases.

In our simulations, we
set the conditions such that
there is no interface kinetics
in direct correlation to 
the conditions imposed in 
the theoretical analysis. 
We derive the same, by setting
the interface relaxation coefficient
$\tau_{\left(\alpha,\beta\right) \gamma}$ through 
a thin-interface analysis \cite{Karma:1996zr,Choudhury:2012uq} 
as,
\begin{align}
\tau_{\left(\alpha,\beta\right) \gamma} &= \dfrac{\left(c^{\left(\alpha,\beta\right)} -
c^\gamma\right)^2}{D^\gamma\dfrac{\partial c^\gamma}{\partial
\mu}}\left(M+F\right)
\end{align}
where $c^{\left(\alpha,\beta\right)}$ are the
equilibrium concentrations
of the phases at equilibrium
at the eutectoid temperature and $M,F$
are solvability integrals derived 
from the thin-interface analysis. The
total sum of $M+F\approx 0.222$ which
is used in the simulations.
As can be seen from the
nature of the phase diagram in Figure~\ref{fig:Phase_Diag},
the difference between these
concentrations change a little
with changing temperature.
Hence, the assumption of using
the values at the eutectoid
temperature holds. 
We state that
 the equation for the
relaxation constant $\tau_{\left(\alpha,\beta\right) \gamma}$
used is strictly valid in cases
pertaining to vanishing diffusivity
in one of the phases, or for instances
of equal diffusivities in both phases.
In the former case, the relation must 
be used with the anti-trapping current which
removes the chemical potential
jump at the interface, 
arising out of artificial solute trapping
as a consequence of choosing a 
thick interface.

The case when the 
diffusivities in both the phases is arbitrary,
the problem is not as trivial. It is to be 
noted that the whole class 
of problems which falls in the category of two-phase transport 
through a complex structure inherently exhibit 
thin-interface effects for stationary/moving
interfaces. Despite some recent progress,
so far no method has been found to 
completely eliminate thin-interface effects 
in the case of arbitrary diffusion coefficients 
in the two phases with a non-stationary interface
\cite{Almgren:1999kx,Nicoli:2011vn,Plapp:2011kx,Steinbach:2009uq}
in a closed form manner.
One of the principal reasons  
for non-closure of the problem,
is that the artificial discontinuous jump effects (arising
out of arbitrary diffusivities) are independent of the 
velocity of the interface.
This implies that 
such effects cannot be removed through the imposition
of an anti-trapping current that has been previously used
for removing the artificial solute trapping effects
in one-sided diffusion problems. While few hints to the
solution by the introduction of tensorial mobilities
\cite{Nicoli:2011vn} and through the usage of 
artificial parameter $\chi$ (related to diffusive current in solid)
\cite{Ohno:2009uq}
are present, a closed form solution is still 
in the process of being worked out. This in 
all certainty is not the highlight or aim of
the present work.

The reason which allows us to
take the liberty of 
overlooking these defects and derive 
meaningful results, is that the interface
width used in the present problem
is of a very small magnitude. 
It is worth noting that the interface width is
proportional to the capillary length of 
the phases and in the present formulation, 
scales inversely with the
factor $\dfrac{\partial^2 f}{\partial c^2}$.
Apparently, while the capillary
length of the phases, austenite and ferrite
are close to each other, the case of cementite
is very restrictive to the choice of the 
interface width. By deriving
leverage out of the fact that
the thin-interface defects scale with the
interface width, a small choice of the 
interface width in present simulations 
allows us to limit the magnitude
of the thin-interface defects, and facilitates 
reasonably quantitative results.

In order to derive the relaxation constant, 
we utilize the mobility of the austenite 
and assume that even though the gradients 
of the chemical potential exist in the solid, 
the principal driving force is still due to the gradient
of the chemical potential in the austenite.
In the absence of a closed form solution, 
this seems like a reasonable choice. 
We substantiate our claim in the following sections,
that the errors due to this assumption, 
do not seem to affect the 
quantitative aspects of our results.

\section{Comparison between theory and simulation}
\label{sec:comparison}
In this section, we compare the growth velocity 
of pearlite obtained from phase-field simulation 
with the analytical results 
in the two regimes: diffusion in austenite (only) 
and diffusion in austenite as well as ferrite. 
The simulation set-up 
comprises of a bounding box with periodic 
boundary conditions in the transverse direction, 
while no flux boundary conditions are used in 
the growth direction. The bounding box width 
in the transverse direction directly controls 
the lamellar spacing $\lambda$ such that 
the pearlitic composition is retained 
(88\% ferrite and 12\% cementite). 
The chemical potential in the bounding 
box is initialized with the equilibrium 
chemical potential for eutectoid
 transformation. The box width in the 
 growth direction is chosen 10
 times larger than the lamellar width
 such that the chemical potential gradient 
 in bulk remains uniform at successive simulation 
 time-steps. In order to completely rule out
 the possibility of non-uniformity of chemical potential 
 in bulk phase, the simulation is carried in a moving frame 
 (also known as shifting-box simulation). 
 In the present simulations, the domain is shifted 
in the growth direction (upwards) by 
 adding a row of grid-point at the top of domain
 and discarding off a row of grid-points at the 
 bottom, every time the advancing lamellar 
 front fills up 10\% of the simulation box.
Further, we have also ensured
 that the 10\% of the box, comprising of the
 growing phases, have sufficient number of cells
 to describe the diffusional field. Although, 
 we cannot claim to be error proof, a good match
 with analytical results, confirms, that the 
 deviations are not large enough to influence 
 the results obtained.
 To find the co-operative lamellar growth rate, the
 previously discarded grid-points are
 aggregated back and the 
 position of advancing interface is determined
 by finding the position of the contour line
 $\phi_{\alpha}-\phi_{\beta} = 0$ through a linear 
 interpolation of the neighboring values (where
 the sign of $\phi_\alpha - \phi_\beta$ changes).
In the present context, $\alpha$ and $\beta$ denote any two phases,
 between which the interface is to be isolated.
 
\begin{figure}
\begin{center}
\includegraphics[scale=0.9]{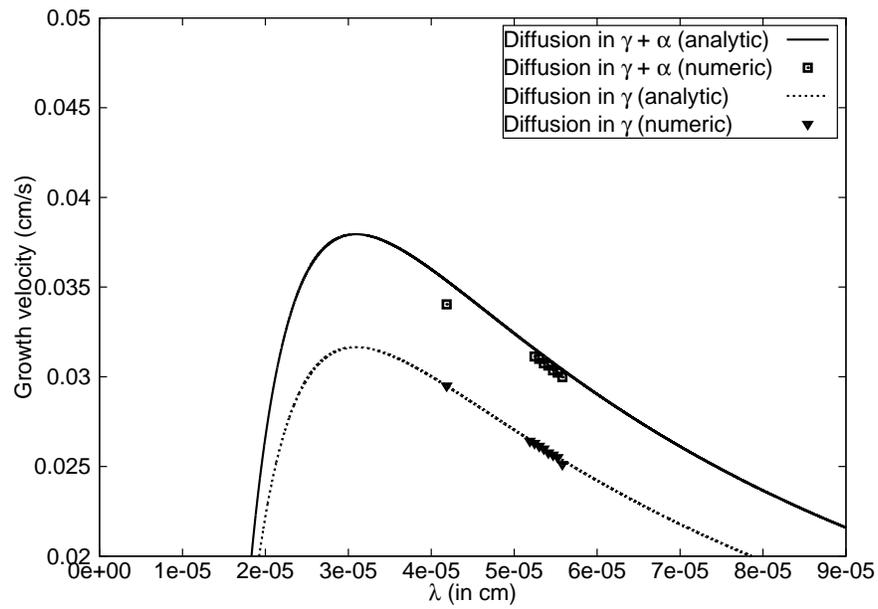}
\caption{Comparison of pearlitic growth front velocities as a function of
lamellar spacing at constant undercooling $\left(\Delta T = 10 K\right)$ derived
from a Jackson-Hunt-type calculation with phase-field results in respective
diffusion regimes.}
\label{fig:comp_anly_num}
\end{center}
\end{figure}

 The rate of change of the position of the interface
 in transverse direction is plotted as a function of time,
 and the simulation is run until there is no more
 change in the velocity of the interface,
 which indicates the steady-state
 has been attained. The procedure described above 
 is repeated to calculate the steady state velocities 
 for different lamellar widths '$\lambda$' 
and plotted for comparison
 with analytical results as shown in 
Figure \ref{fig:comp_anly_num}.
 
\begin{table}
\caption{Parameters used for analytical calculation and for sharp-interface theory}
\label{tab:input_anly}      % Give a unique label
\centering
%\begin{tabular}{lll}
%\hline\noalign{\smallskip}
\begin{tabular}{l l l}
\hline
Symbol & Value & Units \\
\hline
$T_{E}$ & 989 & K \\ 
$\Delta T$ & 10 & K \\
$\sigma_{\alpha\gamma} = \sigma_{\beta\gamma} = \sigma_{\alpha\beta}$ & $0.49$ & J/$m^{2}$\\ 
$D^{\alpha} = D^{\beta}$ & $2\times10^{-9}$ & $m^{2}$/s\\ 
$D^{\gamma}$ & $1\times10^{-9}$ & $m^{2}$/s\\ 
$A^{\alpha,\gamma} = A^{\beta,\gamma}$ & $-1.015385\times10^{-11}$ & $m^{3}$/J\\ 
$A^{\alpha,\beta}$ & $-1.184616\times10^{-12}$ & $m^{3}$/J\\ 
$A^{\beta,\alpha}$ & $-1.9\times10^{-14}$ & $m^{3}$/J\\ 
$c	_{A}^{\alpha}$ & $8.85\times10^{-4}$ & -\\ 
$c	_{A}^{\beta}$ & $0.25$ & -\\ 
$c	_{A}^{\gamma}$ & $0.034433$ & -\\
$c	_{B}^{\alpha}$ & $0.999115$ & -\\
$c	_{B}^{\beta}$ & $0.75$ & -\\
$c	_{B}^{\gamma}$ & $0.965567$ & -\\
$\eta_{\alpha}$ & 0.88 & -\\
$\eta_{\beta}$ & 0.12 & -\\
$\theta_{\alpha\beta}$ = $\theta_{\beta\alpha}$ & $30\degree$ & degrees\\
$\tau_{\alpha\gamma}$ (calculated) & $1.724027\times10^{8}$ & Js/$m^{4}$\\
$\tau_{\beta\gamma}$ (calculated) & $7.118288\times10^{9}$ & Js/$m^{4}$ \\
\noalign{\smallskip}\hline
\end{tabular}
\end{table}

\begin{figure}[!htbp]
\centering
\subfigure[]{\includegraphics[width=0.4\textwidth,height=0.188\textwidth]{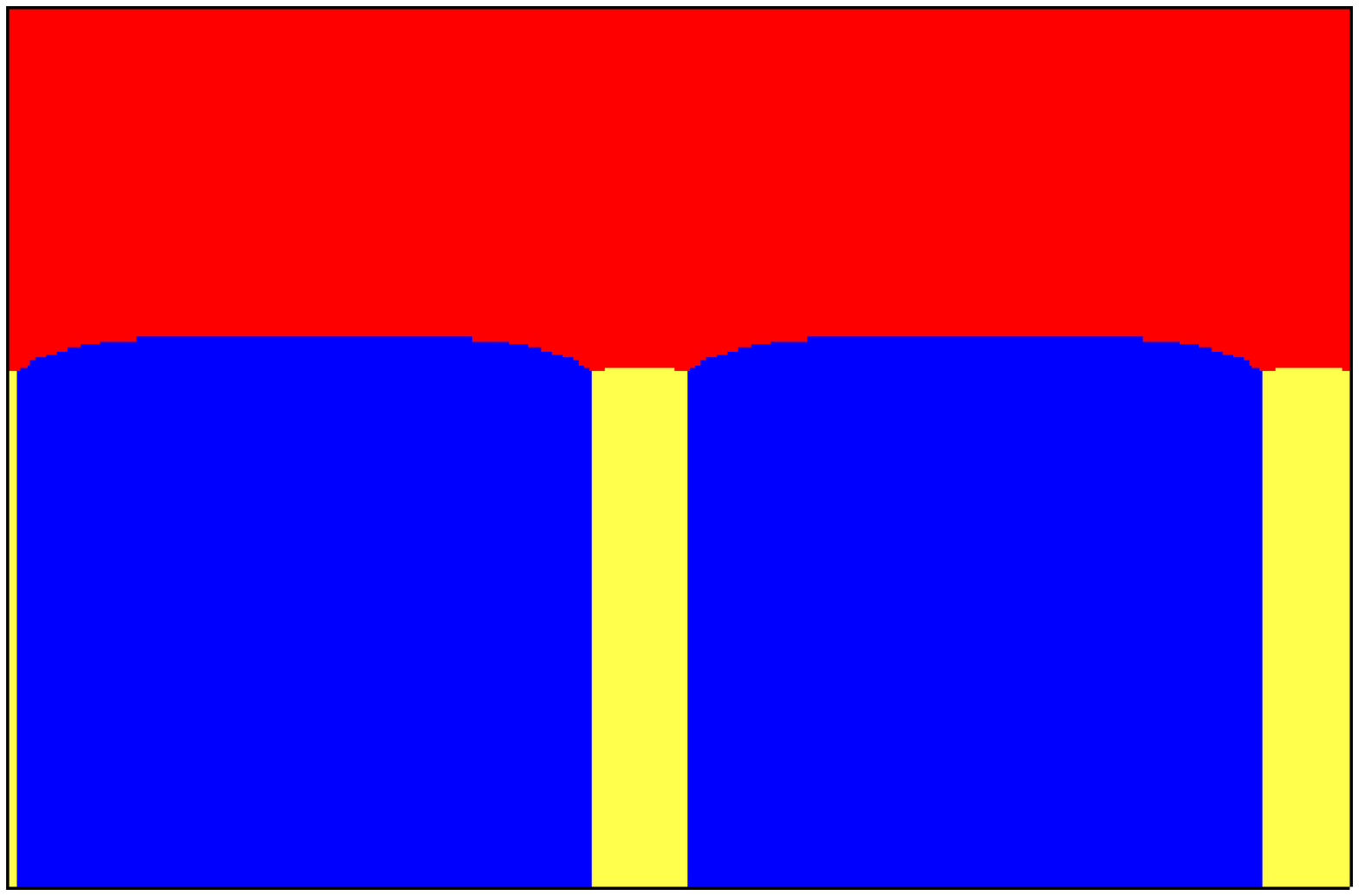}\label{fig:no_diff_phase_alpha}}\qquad
\subfigure[]{\includegraphics[width=0.4\textwidth,height=0.188\textwidth]{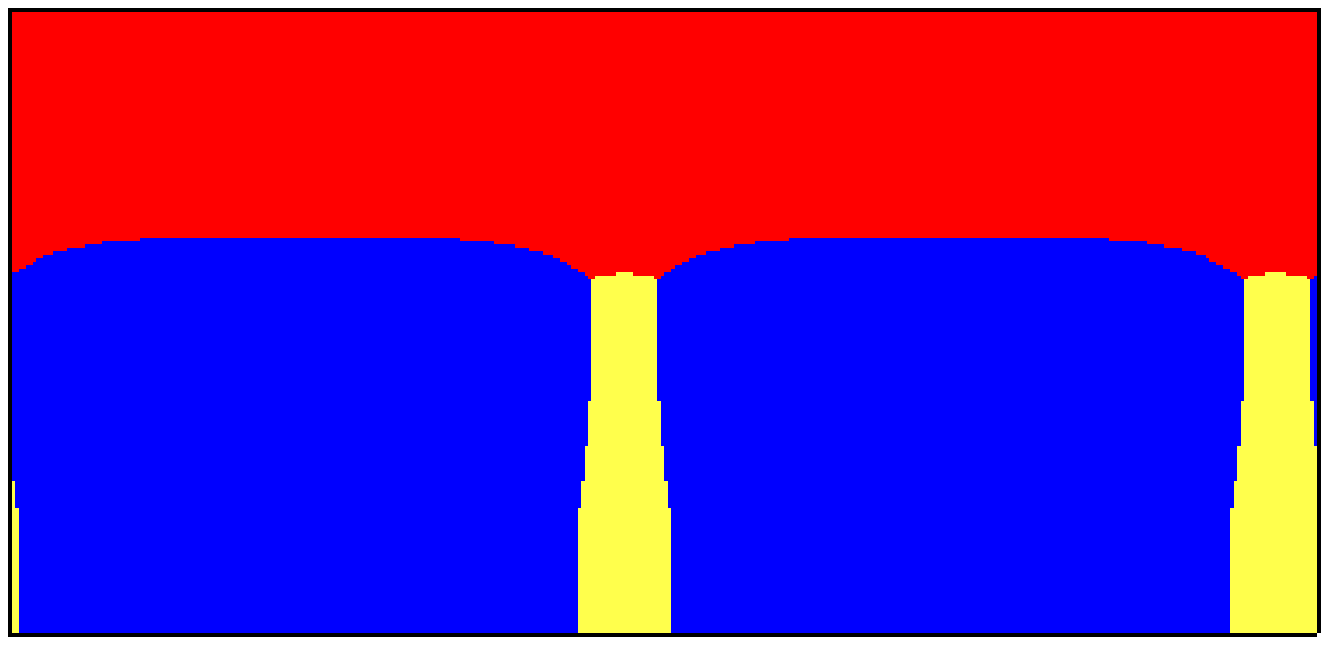}\label{fig:diff_phase_alpha}}\\
\subfigure{\includegraphics[width=0.4\textwidth,height=0.252\textwidth]{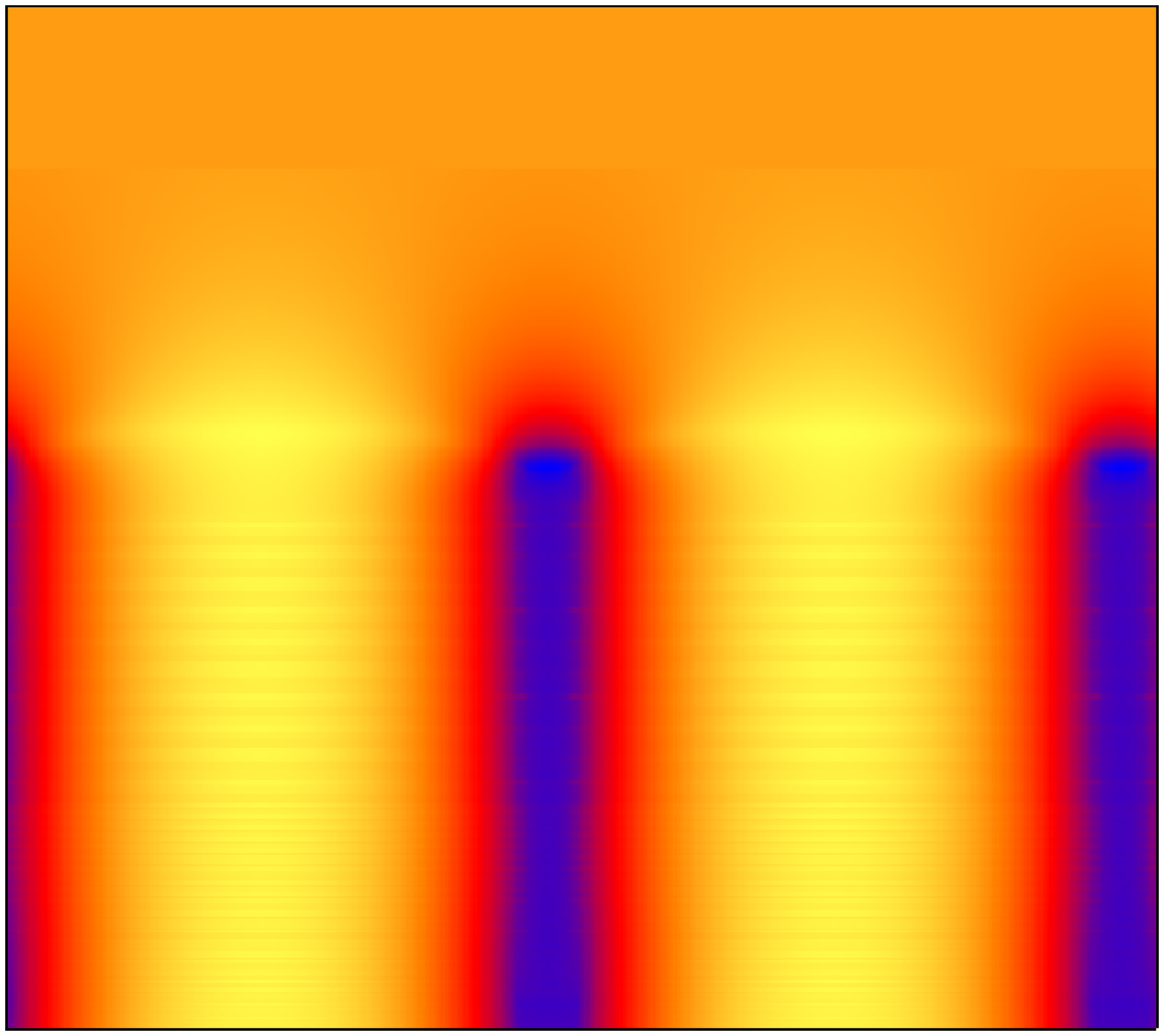}\label{fig:no_diff_conc_alpha}}\qquad
\subfigure{\includegraphics[width=0.4\textwidth,height=0.252\textwidth]{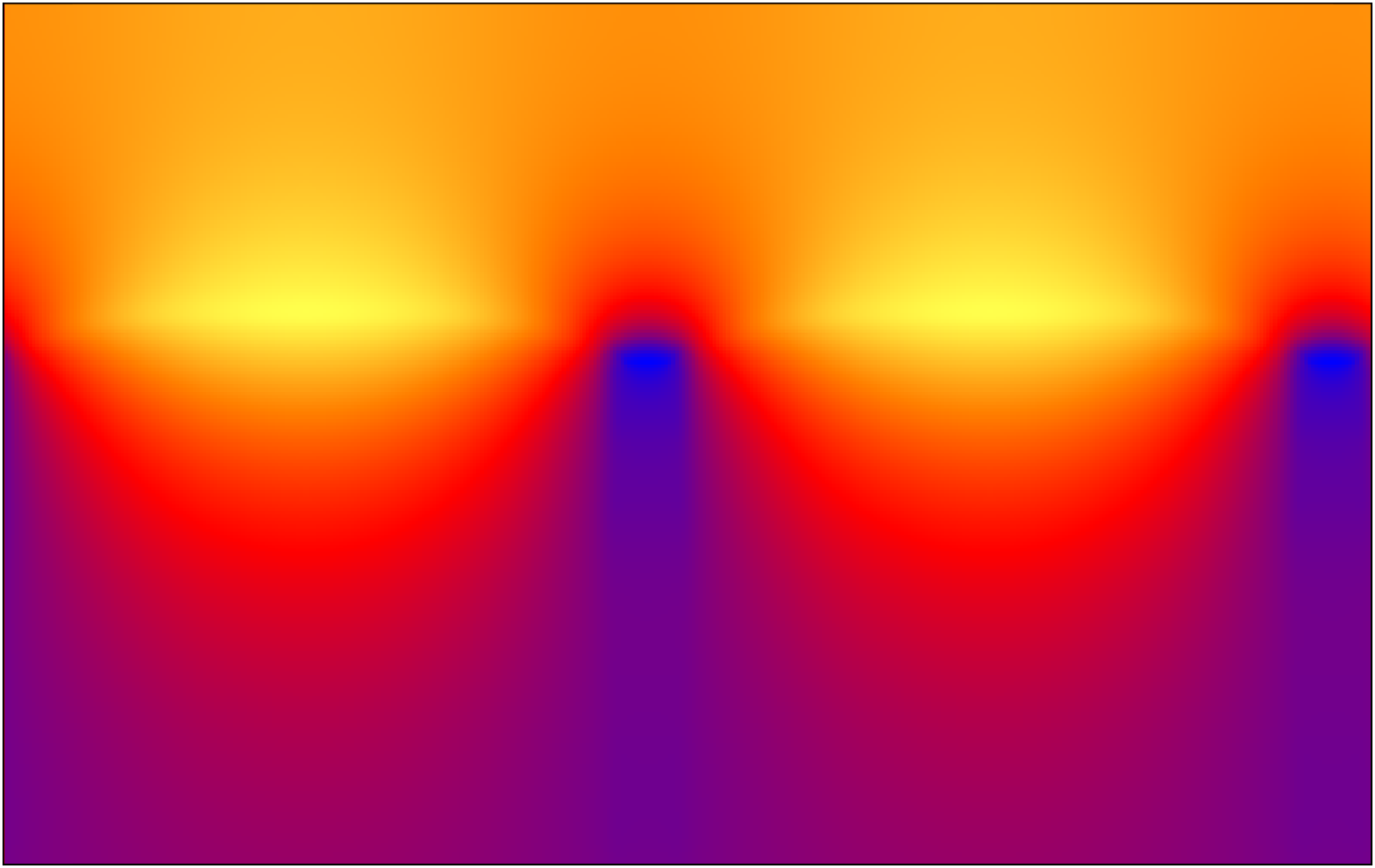}\label{fig:diff_conc_alpha}}\qquad
\setcounter{subfigure}{2}
\subfigure[]{\includegraphics[width=0.4\textwidth]{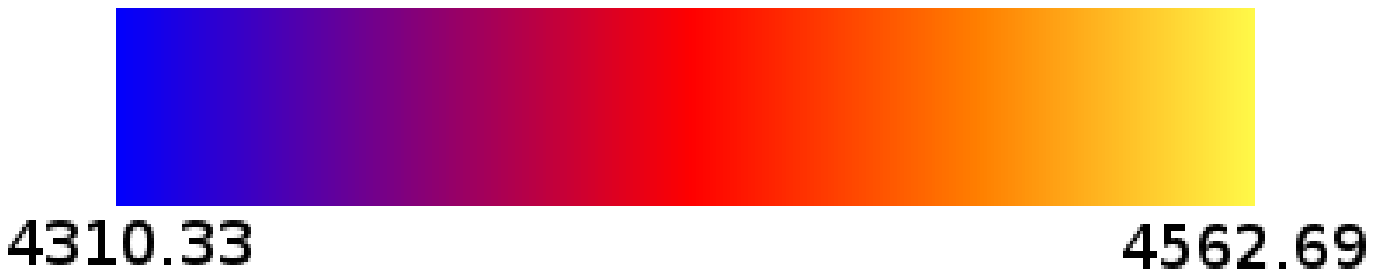}\label{fig:colormap_no_diff_conc_alpha}}\qquad
\subfigure[]{\includegraphics[width=0.4\textwidth]{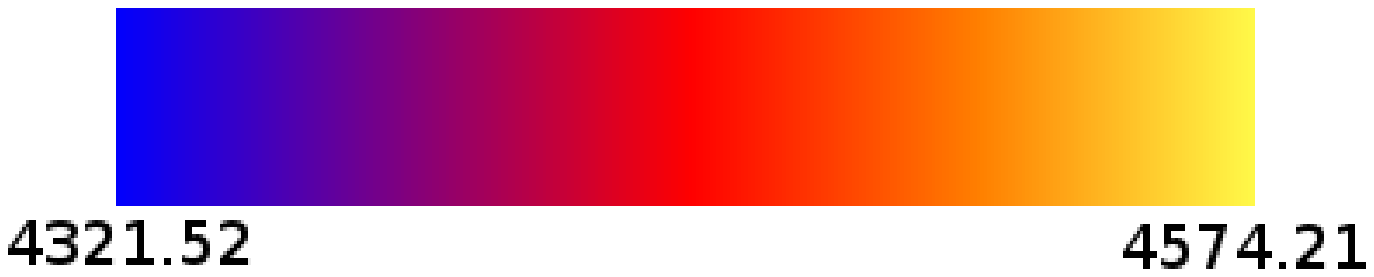}\label{fig:colormap_diff_conc_alpha}}\qquad\\
\caption{Phase patterns (austenite in red, ferrite in blue and cementite in yellow) and chemical potentials plotted for corresponding cases: (a) and (c) Diffusion in austenite (only) and (b) and (d) Diffusion in austenite as well as ferrite. It is to be noted that the pictures above are merely snapshots 
depicting the region of interest and not the entire simulation box itself. The pearlite lamella width is 200 $\mu$m which is half of the illustrated box-length.
\label{fig:phase_conc_comp}}
\end{figure}

The parameters used for analytical calculation 
of the lamellar growth velocity as well as 
for sharp-interface theory as summarized 
in Table~\ref{tab:input_anly}. 
The phase profile obtained from the phase-field simulation 
for the two cases: diffusion in austenite (only) and diffusion 
in austenite as well as ferrite are shown in 
Figure~\ref{fig:no_diff_phase_alpha} and 
Figure~\ref{fig:diff_phase_alpha} respectively. 
The corresponding plots for chemical potential are shown in 
Figure~\ref{fig:no_diff_conc_alpha} 
and Figure~\ref{fig:diff_conc_alpha}. 
It is to be noted that the phase-field simulation 
pictures shown in fig. \ref{fig:phase_conc_comp}
are merely snapshots of the region of interest
and do not represent the
entire simulation box.
The growth velocities of pearlite evolving in 
austenite at an undercooling of 10K in both 
the regimes are plotted in 
Figure~\ref{fig:comp_anly_num}. 
We observe a reasonable overlap of
analytically predicted growth velocity 
with phase-field results in absence
of diffusion in ferrite. However, a small 
deviation is observed near the 
critical lamellar spacing 
when diffusion in ferrite is also 
accounted for along with diffusion 
in austenite, while 
we observe tapering of cementite near the 
growth front due to diffusion in ferrite. 
It is noteworthy, that such a taper 
causes a modification of the triple point
orientation with respect to the growth 
direction, and thereby the triple point
angles assumed in the analytical derivations 
differs from those resulting in 
the simulations. This indeed 
causes modifications in the velocities
near to the critical spacing and the
critical spacing itself, which is also
reflected in our simulations. However, 
for spacings further ahead, where 
the capillary term is less dominant
with respect to the solutal effect,
the agreement is better. Further, 
as already shown in previous
works \cite{Folch:2005uq}, there
exists a difference between 
theoretical analysis of 
the Jackson-Hunt type
and quantitative phase-field 
simulations, which arises because
of the inherent assumptions used
in the analytical derivations.

\pagebreak
\section{Concluding remarks}
In the present work, we generalize the Jackson-Hunt
analysis in order to incorporate diffusion in 
the growing phases, in addition to the diffusion
in the bulk. The results of the analysis predict
that the growth front velocities change
by a factor $\rho$, with respect to the case
where diffusivity is absent in the growing phases.
Phase-field simulations, conducted with 
a thermodynamically consistent model, 
confirm the theoretical predictions fairly
well. In addition, the simulations predict 
the morphology of the growth front showing a 
tapering of cementite in the direction of growth. 

It is worth clarifying that the present work
does not aim to represent the eutectoid transformation
in Fe-C systems, \textit{in totality}. 
Our studies are limited to \textit{only} a part of 
the entire eutectoid transformation 
phenomena, which is to analytically and numerically
investigate the influence of
diffusion in the growing phases, on the lamellar
front velocity, which has been adequately achieved 
through our results.
The presence of a stoichiometric
phase like cementite imposes a limitation on the choice of
length scale and therefore the interface width, 
thereby limiting the thin-interface defects arising
due to having arbitrary diffusivities in all phases.
This argument is further accentuated by a good agreement 
of analytical and numerical results as plotted in 
fig. \ref{fig:comp_anly_num}. Further, this 
near-overlap also suggest that the aim of the
present work has been adequately accomplished.
Thus, the present phase-field simulations
decomposes the effect of dual-diffusion 
mode (in bulk as well as growing phases), 
in isolation i.e. without considering
additional effects for e.g. the 
lattice strains \cite{Steinbach:2007vn}
which influence the growth morphology. 

The complete problem of  
eutectoid transformation however, is complicated and a precise description
is not the aim of the present work.
The additional ingredients, include the contribution from grain-boundary
which is an interesting direction for
future research. Therefore, for 
comparison with experiments and particularly, 
to present a detailed argumentation with
respect to the previous findings 
\cite{Nakajima:2006ve, Nakajima:2007fk}
based on the results of this article would
not be meaningful.
To arrive at a reasonable overlap between 
simulations and experiments, it is 
necessary to consider the combined influence 
of bulk and grain-boundary diffusion
together with transformation strains
in the present phase-field model.
We hope to report on an all-inclusive 
approach for modeling
lamellar growth of pearlite, in the near future.

It will also be 
intriguing to address the faster kinetics of 
re-austenisation or pearlite dissolution (as 
compared to pearlite growth) by deriving 
and transferring ideas from the present work. 
To this end, 
studies need to be 
conducted to establish the
exact mechanism and primary
diffusion regime that
governs the kinetics of such transformations. 
Low undercooling and presence 
of pre-existing cementite particles in austenite, 
alter the 
pearlite growth morphology from 
lamellar to spherodized
widely known as "divorced pearlite" 
\cite{Verhoeven:1998fk}. The phenomena need to be
investigated as simulations contribute in gaining 
further insight on spherodizing
behavior or non co-operative growth of 
pearlite during widely used annealing treatment of steel.
\label{sec:conclusion}

\section*{Acknowledgements}
KA, CQ, SS and BN acknowledge the financial support from the German Research Foundation (DFG) in the framework of Graduate School - 1483. KA also thanks Center for Computing and Communication  at RWTH Aachen University (HPC Cluster) for computational resources.

% end of file template.tex



%% References


\bibliographystyle{model3-num-names}
\bibliography{pearlite}






% \end{thebibliography}

\end{document}